\documentclass[journal,twoside,web]{ieeecolor}
\usepackage{generic}
\usepackage{cite}
\usepackage{amsmath,amssymb,amsfonts}
\usepackage{algorithmic}
\usepackage{graphicx}
\usepackage{textcomp}
\usepackage{siunitx}
\usepackage{upgreek}
\usepackage{soul}

\def\BibTeX{{\rm B\kern-.05em{\sc i\kern-.025em b}\kern-.08em
    T\kern-.1667em\lower.7ex\hbox{E}\kern-.125emX}}
\begin{document}
\title{Cryogenic MOS Transistor Model}
\author{Arnout Beckers, Farzan Jazaeri, and Christian Enz
\thanks{This project has received funding from the European Union's Horizon 2020 Research \& Innovation Programme under grant agreement No. 688539 MOS-Quito (MOS-based Quantum Information Technology) which aims to bring quantum computing to a CMOS platform.}
\thanks{The authors are with the Integrated Circuits Laboratory (ICLAB) at the Ecole Polytechnique F\'ed\'erale de Lausanne (EPFL), 2000 Neuch\^atel, Switzerland
(e-mail: arnout.beckers@epfl.ch).}
}
\maketitle
\begin{abstract}
This paper presents a physics-based analytical model for the MOS transistor operating continuously from room temperature down to liquid-helium temperature (4.2\,K) from depletion to strong inversion and in the linear and saturation regimes. The model is developed relying on the 1D Poisson equation and the drift-diffusion transport mechanism. The validity of the Maxwell-Boltzmann approximation is demonstrated in the limit to zero Kelvin as a result of dopant freeze-out in cryogenic equilibrium. Explicit MOS transistor expressions are then derived including incomplete dopant-ionization, bandgap widening, mobility reduction, and interface charge traps. The temperature-dependency of the interface trapping process explains the discrepancy between the measured value of the subthreshold swing and the thermal limit at deep-cryogenic temperatures. The accuracy of the developed model is validated by experimental results on a commercially available 28 nm bulk CMOS process. The proposed model provides the core expressions for the development of physically-accurate compact models dedicated to low-temperature CMOS circuit simulation.
\end{abstract}


\section{Introduction}
\label{sec:introduction}
\IEEEPARstart{A}{dvanced} CMOS processes perform increasingly well from room temperature down to deep-cryogenic temperatures ($<$ 10 K)\cite{balestra_physics_2017,flandre,essderc,jeds}. At these temperatures the ideal switch with a step-like subthreshold slope comes within reach\cite{rogers1968most}. Furthermore, cryo-electronics\cite{kirschman1985cold,gutierrez2000low,balestra_device_2001} can provide an interface with superconducting devices on the quest for exascale supercomputing\cite{holmes_energy-efficient_2013}. Ultimately, quantum-engineered devices controlled by cryo-CMOS circuits can bring new functionality to existing computing technologies\cite{zwanenburg_silicon_2013,de_franceschi_hybrid_2010}.  

Large-scale integration of silicon spin qubits\cite{pla2012single,maurand2016cmos} and cryo-CMOS control circuits is envisioned to take solid-state quantum computing to the next level\cite{vandersypen_interfacing_2017}. Digital, analog, and RF
CMOS circuits\cite{ekanayake_characterization_2010,reilly_engineering_2015,rahman_cryogenic_2016} are then required to operate at millikelvin temperatures for initialization, manipulation, and read-out of the qubits, as well as error correction\cite{divincenzo_physical_2000,nielsen_quantum_2010}. Since the cooling power at millikelvin temperatures is reduced, the system could feature a cryogenic temperature gradient, where the control circuits operate at a higher cryogenic temperature than the qubits,\,e.g.\,4.2\,K\cite{ekanayake_characterization_2010}. However, the optimal design of power-hungry and thermal-noise dissipating circuits operating in close proximity to the qubits is yet to be explored. In this context, the main hurdle to overcome is the lack of compact MOS transistor models in circuit simulators, remaining physically accurate below 10 K\cite{ekanayake_characterization_2010,rahman_cryogenic_2016}.  
\section{Cryo-MOS Transistor Modeling}
The low-temperature circuits developed for spacecraft\cite{reveret2014cesar,ybecreten}, scientific equipment\cite{hoff_cryogenic_2015}, ultra-low-noise detectors\cite{okcan_cryogenic_2010}, cryobiology\cite{ihmig_frozen_2013}, and others, have been custom-designed relying on a semi-empirical approach. This approach requires laborious and expensive low-temperature measurements to extract model parameters for tuning room-temperature compact models to the target low-temperature \cite{okcan_cryogenic_2010,martin_ekv3_2009,creten_cryogenic_2007}. Empirical temperature-scaling laws have been added to the room temperature physics-based MOS transistor model\cite{tsividis2011operation,sze2006physics} to capture cryogenic operation down to 4.2 K\cite{akturk2010compact,akturk2007device,akturk_compact_2010}. However, the discrepancy between the measured value of the subthreshold swing for a long device at 4.2\,K ($\approx$\,10\,mV/decade)\cite{essderc,beckers_eurosoi,jeds}, and the theoretical thermal limit, $U_T\ln 10$ ($\approx$\,0.8 mV/decade) reveals that something more fundamental is missing. As we will demonstrate along this paper, important physical phenomena at low temperatures such as interface trapping\cite{hafez_assessment_1990,sze2006physics} and incomplete ionization\cite{jonscher1964semiconductors,foty1990impurity} have not been properly included to date. Furthermore, the intrinsic carrier concentration, $n_i$, takes on extremely small values below \SI{10}{K}, causing arithmetic underflow in implemented analytical expressions or convergence problems in computer-aided-design simulations\cite{jaeger1980simulation,turowski2012device,kantner2016numerical}. Therefore, standard references on semiconductor devices treat only the cryogenic equilibrium condition in bulk semiconductors above \SI{10}{\K}\cite{sze2006physics,pierret1987advanced,tsividis2011operation}. Analytical device-physics models, starting from the Poisson equation at low temperature, leave a gap unfilled between the zero-Kelvin approximation and \SI{77}{\K}\cite{wu1974mosfet,wilson1986simple,sim1992analytical,hafez_characterization_1990}.

In this work, we develop a MOS transistor model valid from room temperature (RT) down to deep-cryogenic temperatures, entirely based on physics principles and validated with experimental results. We start by verifying the continued validity of the Boltzmann statistics down to the deep-cryogenic regime.
\section{\label{sec:model} MOS Electrostatics from RT to 4.2 K}
We model a long, planar $n$-channel MOS field-effect transistor in silicon, depicted in Fig.\,\ref{fig:mos}. Uniform operation across the width of the transistor is assumed and the gradual channel approximation is adopted. The electrostatics can then be described by the 1D Poisson equation \cite{tsividis2011operation,sze2006physics}.  
\subsection{Poisson-Fermi equation}
Merging the 1D Poisson equation with the mobile carrier concentrations, $n$ and $p$, given by Fermi-Dirac statistics, gives
\begin{equation}\label{MBFermipoisson}
\begin{split}
\frac{\partial^2\psi(y)}{\partial y^2}&=-\frac{q}{\varepsilon_{si}}\left(-n+p-N_A^-\right), 
\end{split}
\end{equation} 
where $q$ is the elementary charge, $\varepsilon_{si}$ the silicon permittivity, and $\psi\triangleq(E_F-E_i)/q$ the potential, with $E_F$ the Fermi-level and $E_i$ the intrinsic energy level. The first term on the RHS of equation (\ref{MBFermipoisson}) represents the electron contribution, $n$, the second term the hole contribution, $p$, and the third term the ionized dopant-contribution, $N_A^-$. \subsubsection{Incompletely-ionized dopants}
\begin{figure}[t]
	\centering
	\includegraphics[scale=0.65]{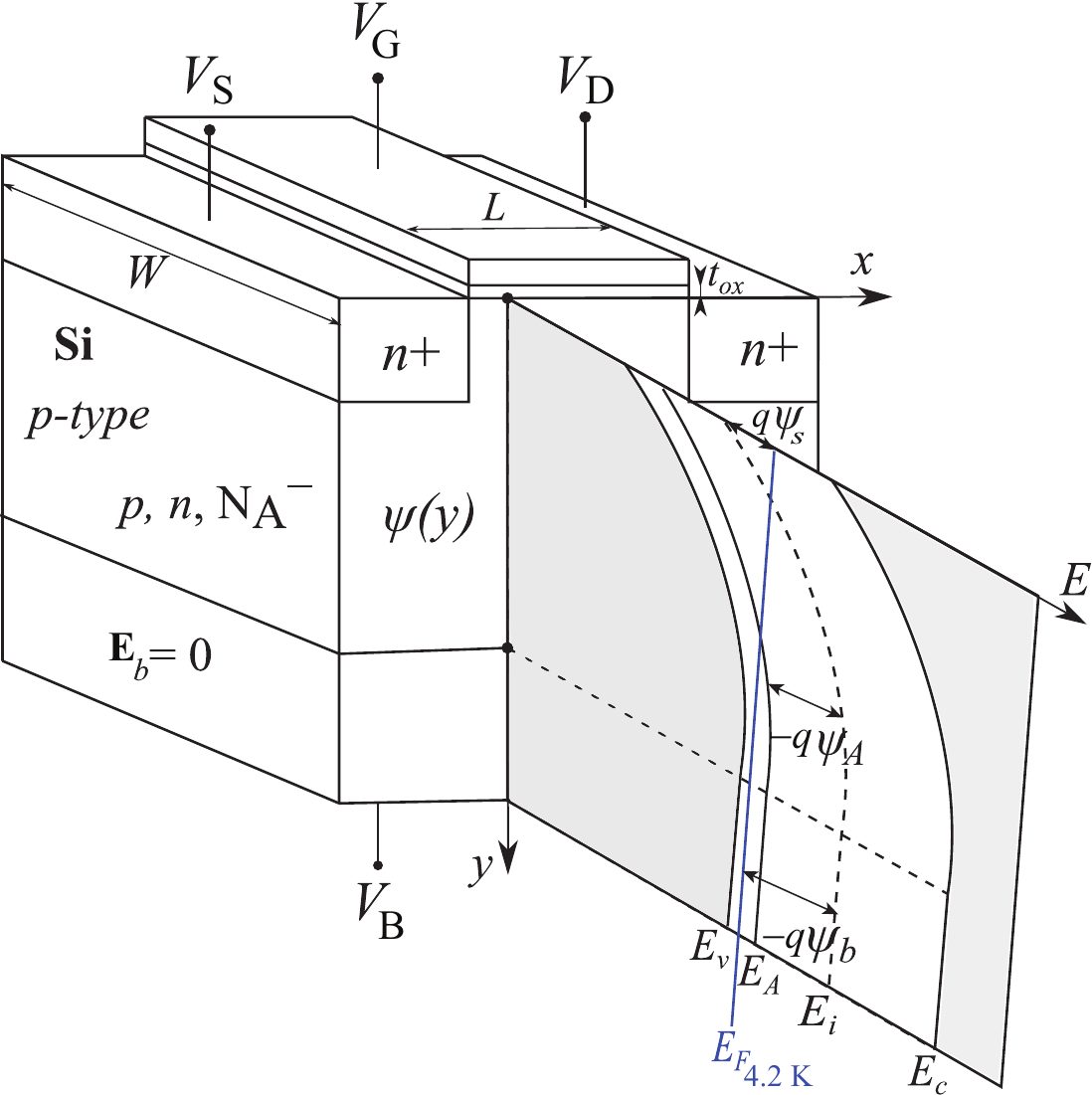}
	\caption{\label{fig:mos}Schematic representation of a long $n$MOS transistor with annotated band diagram. The drift-diffusion and Poisson equations are solved along the $x$ and $y$-directions respectively. At 4.2\,K, and for $N_A=\SI{e18}{\per\centi\meter\cubed}$, $E_F$ lies below $E_A$ in the bulk (see Fig.\,\ref{fig:extrinsic}a), leading to bulk freeze-out according to (\ref{inco}). When $E_A$ bends under $E_F$ near the surface, the acceptor dopants become rapidly completely ionized due to field-assisted ionization. The quasi-Fermi potential is not taken into account in this figure.}
\end{figure}

Under thermal equilibrium, both at room and cryogenic temperatures, the majority carrier concentration can defer from the implanted doping value, $N_A$, due to incomplete ionization of the dopants. In cryogenic equilibrium, incomplete ionization is strong and known as freeze-out, since thermal dopant-ionization is very low\cite{foty1990impurity}. However, during MOS operation, also field-assisted ionization comes into play. Fermi-Dirac statistics provides a fundamental way to model incomplete ionization which includes both dopant-ionization mechanisms. The concentration of ionized dopants, $N_A^-$, is then equal to the total concentration of implanted dopants times the Fermi-Dirac occupation probability of the acceptor energy $E_A$, i.e.\,$N_A\times f(E_A)$, or
\begin{equation}\label{inco}
N_A^-=\frac{N_A}{1+g_Ae^{\frac{E_A-E_{F,n}}{kT}}}=\frac{N_A}{1+g_Ae^{\frac{\psi_A-(\psi-V_{ch})}{U_T}}}, 
\end{equation}
where the electron quasi-Fermi-level is given by $E_{F,n}=E_F-qV_{ch}$. The RHS of (\ref{inco}) is obtained by replacing $E_A-E_{F,n}$ with $E_A-E_i+E_i-E_{F,n}$ in the exponential term, and by defining an acceptor potential, $\psi_A\triangleq(E_A-E_i)/q$, as indicated in Fig.\,\ref{fig:mos}. The channel voltage, $V_{ch}$, denotes the shift of the quasi-Fermi-potential due to the drain-to-source voltage, $V_{\mathrm{DS}}$. The second expression in (\ref{inco}) highlights the two dopant-ionization contributions, i.e.\, the potential (field-assisted ionization\cite{foty1990impurity}) and temperature (thermal ionization). The acceptor-site degeneracy factor, $g_A$, is set to four due to fourfold degeneracy (heavy-light hole, spin up-down)\cite{pierret1987advanced,sze2006physics}. Note that setting $g_A$ to zero is equivalent to assuming complete ionization.

\subsubsection{Mobile carrier concentrations}Since $n$ and $p$ given by Fermi-Dirac statistics in (\ref{MBFermipoisson}) require numerical integration over energy, this inhibits explicit solutions for the charge densities and current in the MOS transistor. Expressing $n$ and $p$ using Boltzmann statistics allows to obtain such relations. However, the validity of the Maxwell-Boltzmann approximation down to deep-cryogenic temperatures is questionable. It has been reported\cite{kantner2016numerical,sim1992analytical} that semiconductors become strongly degenerate at deep-cryogenic temperatures, preventing its use. This is however inconsistent with the zero-Kelvin limits of the Fermi-level position in the bandgap derived by Pierret\cite{pierret1987advanced}. Therefore, in the next subsection we aim to verify the Maxwell-Boltzmann approximation down to deep-cryogenic temperatures.
\begin{figure*}[t]
	\includegraphics[width=\textwidth]{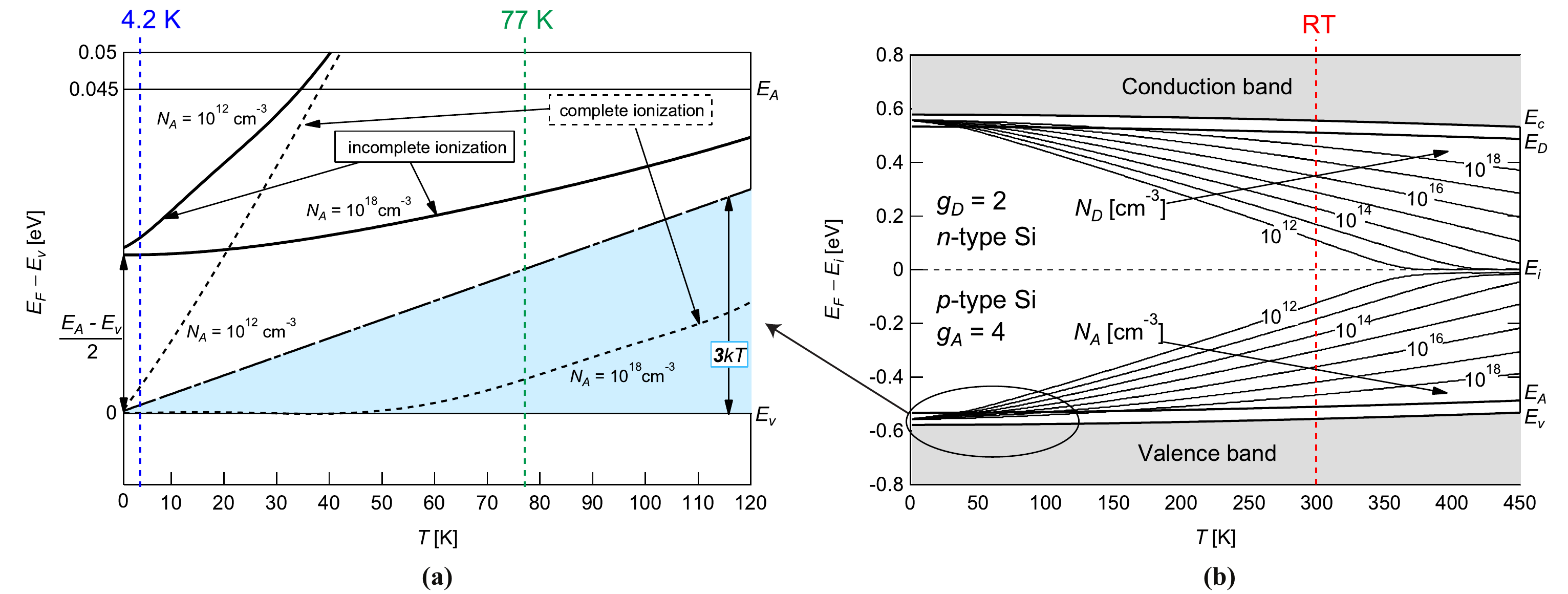}
	\vspace{-0.6cm}
	\caption{Thermal equilibrium in extrinsic bulk silicon. Right: position of the Fermi-level, $E_F$, in the bandgap as a function of doping and temperature. The $E_F$-position is calculated
	from RT down to 100\,mK using an extension of the arithmetic precision and an $E_F$-resolution of 1\,meV. 
	Left: magnified view of the cryogenic regime (below 120\,K). When incomplete ionization is taken into account, the distance of $E_F$ to the valence-band edge, $E_v$, stays larger than $3kT$, validating the use of the Maxwell-Boltzmann approximation down to millikelvin temperatures. This figure applies to the bulk of the MOS transistor in all regions of operation, and to the whole body of the MOS transistor in the flatband condition. Bandgap temperature dependency is taken from Varshni\cite{varshni1967temperature} and a standard, temperature-independent value of $E_A-E_v=0.045$\,eV in Si:B is assumed.}
	\label{fig:extrinsic}
\end{figure*}
\subsubsection{\label{sec:MB}Verification of Boltzmann statistics}
We numerically calculate the position of the equilibrium Fermi-level, $E_{F}$, down to 100\,mK relying on Fermi-Dirac statistics in an extrinsic bulk semiconductor, e.g.\,$p$-type silicon. In this case, the Poisson equation imposes the charge neutrality,~$p_p=N_{A}^-$, where $p_p$ is expressed by Fermi-Dirac statistics\cite{sze2006physics,pierret1987advanced} and $N_A^-$ by (\ref{inco}). This yields an implicit equation for $E_{F}$, which is solved numerically at each temperature and doping value using an extension of the arithmetic precision. As illustrated in Fig.\,\ref{fig:extrinsic}a, below 120\,K, $E_{F}$ remains off the valence band edge with an offset larger than $3kT$ for doping values below the degenerate limit (i.e.\,$N_A\,=\,\SI{4e18}{\per\centi\meter\cubed}$ in Si:B)\cite{sze2006physics,pierret1987advanced,altermatt2006simulationII}. Note that this is predicted correctly only when incomplete ionization is taken into account. Complete ionization\,($g_A=0$)\, would predict an offset smaller than $3kT$ for $N_A\,=\,\SI{e18}{\per\centi\meter\cubed}$, and hence a degenerate semiconductor. It should therefore be emphasized that incomplete ionization maintains the non-degeneracy of a highly-doped semiconductor at temperatures down to 100\,mK. Furthermore, near zero Kelvin, $E_{F}$ tends to saturate at $(E_A-E_v)/2$ for all considered doping values. This corresponds to the zero-Kelvin limit in Pierret\cite{pierret1987advanced} assuming Boltzmann statistics. Using the now validated Maxwell-Boltzmann description for $p_p$, i.e.\,$N_v\exp[(E_v-E_F)/kT]$, in~$p_p=N_{A}^-$, leads to a quadratic equation in $\exp\left[(E_v-E_{F})/kT\right]$ with as solution, 
\begin{equation}\label{limit}
\begin{split}
E_{F}-E_v&=kT\ln\frac{N_v}{N_A}+kT\ln\frac{1+\sqrt{1+4g_A\frac{N_A}{N_v}e^{\frac{E_A-E_v}{kT}}}}{2}. 
\end{split}
\end{equation}
Considering the temperature dependency of $N_v$\cite{sze2006physics,pierret1987advanced}, while taking the limit of (\ref{limit}) to \SI{0}{\K}, leads to $\lim_{T\rightarrow 0 K}E_{F}=E_v+(E_A-E_v)/2$. 

\begin{figure}[]
	\centering
	\includegraphics[width=0.5\textwidth]{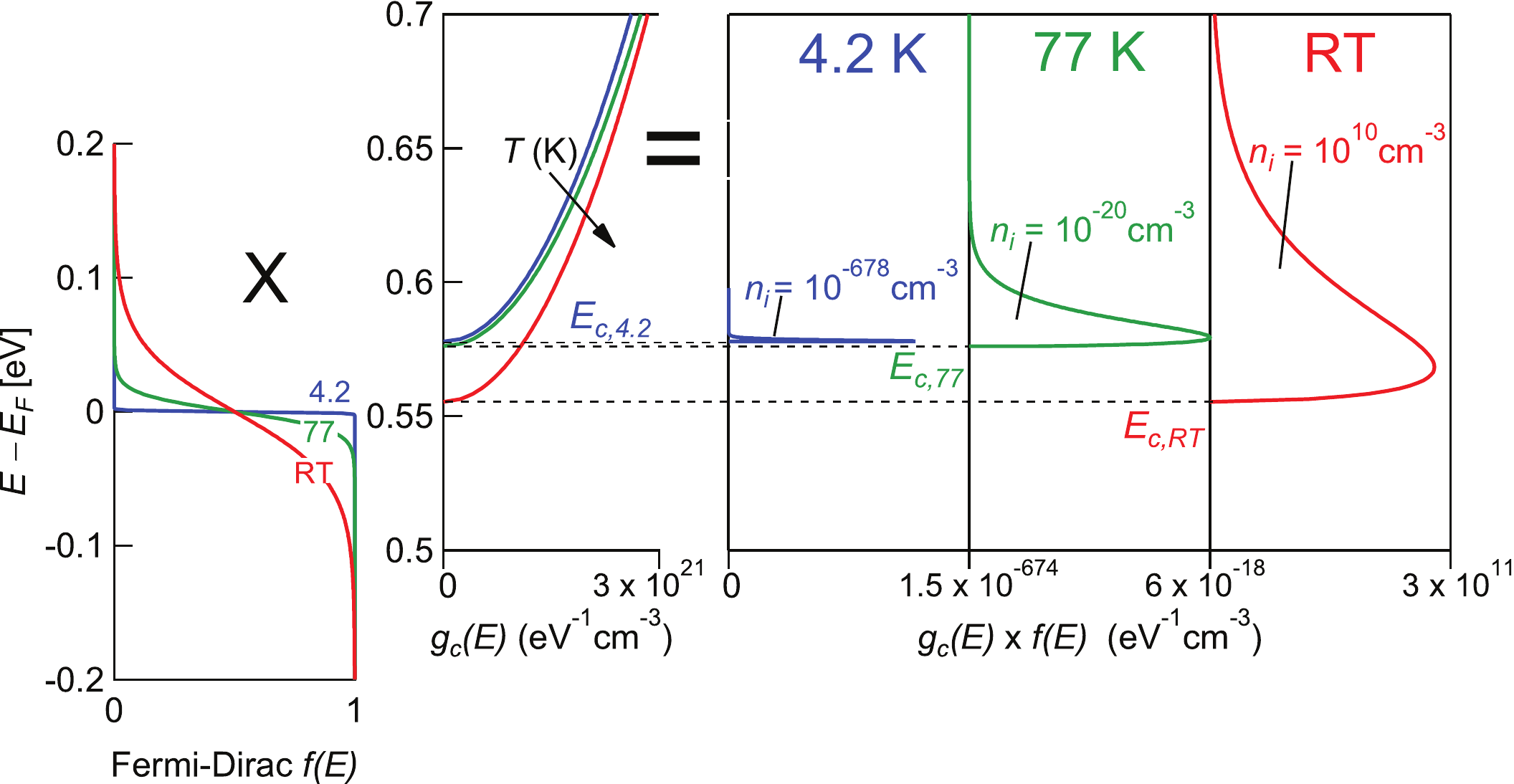}
	\caption{\label{fig:ni}The intrinsic carrier concentration reaches extremely small values at \SI{4.2}{\,K}, left: Fermi-Dirac distribution function approaching a step function at \SI{4.2}{\,K}, middle: density-of-states in the conduction band, right: overlap between the density-of-states in the conduction band and the Fermi-Dirac distribution function at \SI{4.2}{\,K}, \SI{77}{\,K} and room temperature (RT). The overlap function $g_c(E)\times f(E)$ becomes extremely small in magnitude and very peaked at \SI{4.2}{\K}. The area under the overlap function is equal to the intrinsic carrier concentration. Bandgap temperature-dependency used from Varshni\cite{varshni1967temperature} and effective mass values from Pierret\cite{pierret1987advanced}.}	
\end{figure}

Performing the same numerical $E_F$-calculation for an intrinsic semiconductor, the extremely small value of $n_i$ can be verified relying on Fermi-Dirac statistics. The Poisson equation then imposes the charge neutrality, $n=p=n_i$, where $n$ and $p$ are given by Fermi-Dirac statistics. As illustrated in Fig.\,\ref{fig:ni}, this yields $n_i$-values lying outside the range of IEEE double-precision arithmetic ($10^{-308}-10^{308}$), e.g.\,at \SI{4.2}{K}, $\approx$\,\SI{e-678}{\per\centi\meter\cubed}. Therefore, an extension of the arithmetic precision will also be used in the remainder of this work based on Boltzmann statistics, since the carrier concentrations are then expressed through $n_i$.
\subsection{Poisson-Boltzmann equation}
Using the Maxwell-Boltzmann approximation of $n$ and $p$, validated down to deep-cryogenic temperatures in the previous section, we combine the 1D Poisson equation with Boltzmann statistics, which leads to
\begin{equation}\label{MBpoisson}
\begin{split}
\frac{\partial^2\psi(y)}{\partial y^2}&=-\frac{q}{\varepsilon_{si}}\left(-n_ie^{\frac{\psi-V_{ch}}{U_T}}+n_ie^{-\frac{\psi}{U_T}}-N_A^-\right), 
\end{split}
\end{equation} 
where $U_T \triangleq kT/q$ is the thermal voltage. The first term on the RHS of equation (\ref{MBpoisson}) represents the electron contribution, $n$, and the second term the hole contribution, $p$. The intrinsic carrier concentration is given by $n_i=\sqrt{N_cN_v}\exp(-E_g/2kT)$, with $E_g$ the bandgap, and $N_c$ and $N_v$ the effective density-of-states in the conduction band and valence band respectively. The temperature-dependency of $E_g$ as described by Varshni\cite{varshni1967temperature} is used. The extremely small, but finite value of $n_i$ at deep-cryogenic temperatures cannot be assumed zero\textemdash which would be equivalent to the zero-Kelvin approximation\cite{wu1974mosfet} or considering $f(E)$ as a step function\textemdash since this leads to zero mobile carrier concentrations independently of the potential. This is irreconcilable with the observed field-effect and correct functioning of the MOS transistor at 4.2\,K\cite{rogers1968most,essderc}. For $U_T$ small, the exponential factor has a very big dynamic range when $\psi$ changes during MOS transistor operation, large enough to overrule $n_i$ in the multiplication.
\subsubsection{Derivation of  the electric field at the surface}
Introducing (\ref{inco}) for $N_A^-$ in Eq.\,(\ref{MBpoisson}), and then multiplying (\ref{MBpoisson}) on both sides with $2(\partial\psi/\partial y)$ gives
\begin{equation}
\begin{split}
\frac{\partial}{\partial y}\Bigg[\Bigg(\frac{\partial\psi(y)}{\partial y}\Bigg)^2\Bigg]=\frac{2q}{\varepsilon_{si}}\Bigg(&n_ie^{\frac{\psi-V_{ch}}{U_T}}-n_ie^{-\frac{\psi}{U_T}}\\
&+\frac{N_A}{1+g_Ae^{\frac{\psi_A-(\psi-V_{ch})}{U_T}}}\Bigg)\frac{\partial \psi}{\partial y}.  
\end{split}
\label{esintegral}
\end{equation}
Integrating (\ref{esintegral}) from bulk to surface with $\textbf{E}=-\partial \psi/\partial y$ and $\textbf{E}_b=0$ yields
\begin{equation}\label{esint}
\begin{split}
\textbf{E}^2_s=\frac{2q}{\varepsilon_{si}}\int_{\psi_b}^{\psi_s}\Bigg(n_ie^{\frac{\psi-V_{ch}}{U_T}}&-n_ie^{-\frac{\psi}{U_T}}\\&+\frac{N_A}{1+g_Ae^{\frac{\psi_A-(\psi-V_{ch})}{U_T}}}\Bigg)d\psi. 
\end{split}
\end{equation} 
In (\ref{esint}) the additional potential dependence due to field-assisted ionization of the dopants can be straightforwardly integrated as well, i.e. by replacing $N_A$ with $N_A\{1+g_A\exp[(\psi_A-(\psi-V_{ch}))/U_T]-g_A\exp[(\psi_A-(\psi-V_{ch}))/U_T]\}$ in the numerator of the third term and splitting the resulting integral. This gives an expression for the square of the electric field at the surface, 
\begin{equation}\label{es}
\begin{split}
\textbf{E}_s^2 =\frac{2qn_iU_T}{\varepsilon_{si}}&\left(e^{\frac{\psi_s-V_{ch}}{U_T}}-e^{\frac{\psi_b-V_{ch}}{U_T}}+e^{-\frac{\Psi_s}{U_T}}-e^{-\frac{\Psi_b}{U_T}}\right)\\
&+\frac{2qN_A}{\varepsilon_{si}}\Bigg[\psi_s-\psi_b-U_T\ln\frac{f_s(E_A)}{f_b(E_A)}\Bigg],
\end{split}
\end{equation}
where $\psi_b\triangleq (E_{F,b}-E_i)/q$ is the bulk potential and $\psi_s\triangleq (E_{F,s}-E_i)/q$ the surface potential, as indicated in Fig.\ref{fig:mos}. $E_{F,s}$ denotes the Fermi-level at the surface, and $E_{F,b}$ the Fermi-level in the bulk. The logarithmic term in (\ref{es}) is the contribution of incomplete ionization, where 
\begin{equation}\label{fsea}
f_s(E_A)\triangleq\frac{1}{1+g_Ae^{\frac{E_A-E_{F,s}}{kT}}}=\frac{1}{1+g_Ae^{\frac{\psi_A-(\psi_s-V_{ch})}{U_T}}}
\end{equation}
the Fermi-Dirac ionization probability at the surface, and 
\begin{equation}\label{fbEA}
f_b(E_A)\triangleq\frac{1}{1+g_Ae^{\frac{E_A-E_{F,b}}{kT}}}=\frac{1}{1+g_Ae^{\frac{\psi_A-\psi_b}{U_T}}},
\end{equation}
the Fermi-Dirac ionization probability in the bulk, assuming that $V_{ch}$ is zero in the bulk. Both ionization probabilities are qualitatively shown in Fig.\,\ref{fig:mos}. If complete ionization is assumed, then $f_s(E_A)=f_b(E_A)=1$ and the incomplete ionization term cancels in (\ref{es}), leading to the expression widely-used at RT\cite{tsividis2011operation,sze2006physics}. The surface-ionization probability $f_s(E_A)$ is plotted in Fig.\,\ref{fig:ionization} as a function of thermal and field-assisted ionization. Immediately evident is that freeze-out at the surface (arbitrarily defined when $f_s(E_A)<0.2$) is only present when the temperature is below $\approx$~\SI{50}{\kelvin} and the potential is close to the flatband condition ($\psi_s\approx \psi_b$). Above $\psi_b$, the ionization probability rapidly transitions to one due to field-assisted ionization. This transition corresponds to the bending of $E_A$ under $E_{F}$ at the surface in Fig.\,\ref{fig:mos}. Therefore, complete ionization is a valid approximation even at deep-cryogenic temperatures, although the shift in $E_F$ due to incomplete ionization (Fig.\,\ref{fig:extrinsic}a) should be taken into account since it affects the threshold voltage. This $E_F$-shift can be quantified by using $f_b(E_A)$ from (\ref{fbEA}) in the bulk charge neutrality condition, $p_p=N_A^-$, which leads to the quadratic equation $\exp(2\psi_b/U_T)-(n_i/N_A)\exp(\psi_b/U_T)-(g_A/N_A)\exp(\psi_A/U_T)$ with as solution,
\begin{equation}
\psi_b=U_T\ln\frac{n_i}{N_A}+U_T\ln\frac{1+\sqrt{1+4\frac{N_A}{n_i}g_Ae^{\frac{\psi_A}{U_T}}}}{2}. 
\label{psib}
\end{equation}
The second term in (\ref{psib}) is the shift of $E_F$ by including incomplete ionization, which is only dependent on temperature and doping. Assuming complete ionization,\,i.e.\,$g_A=0$, the well-known $U_T\ln (n_i/N_A)$ is obtained. 
\subsubsection{Derivation of the charge densities}                       
\begin{figure*}[t!]
	\includegraphics[width=\textwidth]{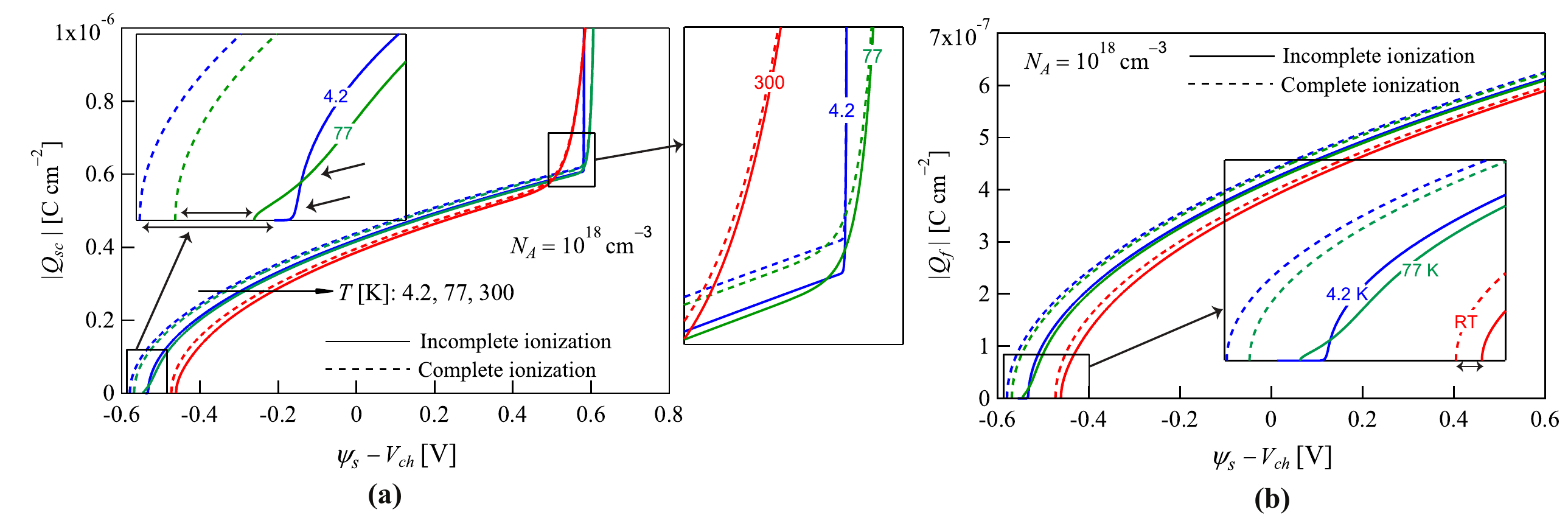}
	\vspace{-0.7cm}
	\caption{\label{fig:Qsc} (a) Total semiconductor charge density, $Q_{sc}$, and (b) fixed charge density, $Q_{f}$, at room temperature (RT, red), liquid-nitrogen temperature (\SI{77}{\kelvin}, green), and liquid-helium temperature (\SI{4.2}{\kelvin}, blue) including incomplete ionization (solid lines) or assuming complete ionization (dashed lines). The potential is swept starting from the bulk potential, $\psi_b$, calculated at a given temperature and doping according to (\ref{psib}). Horizontal arrows show the shifts in $\psi_b$ by including incomplete ionization at a given temperature. Skewed arrows in the insets indicate the kinks at 77 and 4.2\,K due to the transition from incomplete to complete ionization when $E_A$ bends under $E_F$ (Fig.\,\ref{fig:mos}).}	
\end{figure*}

Applying the Gaussian law over the semiconductor body in Fig.\,\ref{fig:mos}, the total semiconductor charge density per unit area, $Q_{sc}$, is obtained by $Q_{sc}=-\varepsilon_{si}\textbf{E}_s$, with $\textbf{E}_s$ given by (\ref{es}). The obtained $Q_{sc}$ is plotted in Fig.\,\ref{fig:Qsc}a at RT, \SI{77}{\kelvin}, and \SI{4.2}{\kelvin}. For 77\,K and 4.2\,K, small kinks are noticeable close to $\psi_b$ due to the transition from incomplete to complete ionization when $E_A$ bends under $E_F$ at the surface ($E_{F,s}$), or equivalently, $\psi_s$ becomes less negative than $\psi_A$. Above this transition, $f_s(E_A)\approx1$ according to (\ref{inco}). At RT, $E_{F}$ lies above $E_A$ in the flatband condition (see Fig.\ref{fig:extrinsic}b) and hence no transitional kink is noticeable. There is however a $\psi_b$-shift also at RT due to incomplete ionization according to (\ref{psib}). Note that for complete ionization (dashed lines) no kinks are observed since the logarithmic term cancels in (\ref{es}). Assuming the charge-sheet and fully-depletion approximations\cite{tsividis2011operation}, the fixed charge density per unit area, $Q_f$, is given by  
\begin{equation}\label{qf}
Q_f=-\varepsilon_{si}\sqrt{\frac{2qN_A}{\varepsilon_{si}}\big(\psi_s-\psi_b\big)-\frac{2qN_AU_T}{\varepsilon_{si}}\ln\frac{f_s(E_A)}{f_b(E_A)}}.
\end{equation}
Relying on the charge neutrality, the mobile charge density per unit area, $Q_m$, can be obtained from $Q_m = Q_{sc}-Q_f$, resulting in Eq.(\ref{qm}). $Q_m$ is plotted in Fig.\,\ref{fig:Qm}a for RT, \SI{77}{\kelvin}, and \SI{4.2}{\kelvin}. As can be observed in this figure, incomplete ionization does not affect the turn-on rate of $Q_m$, but contributes a small decrease in the charge threshold voltage, due to the shift of the $E_F$-position closer towards the conduction-band edge as shown in Fig.\,\ref{fig:extrinsic}a and derived in (\ref{psib}).

Therefore, from this section we conclude that incomplete ionization cannot explain the offset between the measured subthreshold swing at 4.2\,K and the thermal limit. As we will show in the next section, the temperature-dependent occupation of interface charge traps can degrade the subthreshold swing down to 4.2\,K. 
\subsubsection{\label{sec:traps}Interface charge traps}
Defects and lattice breaking at the oxide-semiconductor interface introduce trap energy levels, $E_t$, in the bandgap which degrade the control of the gate-to-bulk voltage, $V_{\mathrm{GB}}$, over the channel. In what follows, the Fermi-Dirac occupation of interface traps, $f(E_t)$, is included in the surface-boundary condition and the effect on the $Q_m$ turn-on rate is analyzed at 4.2\,K. 
\begin{figure}[b!]
	\centering
	\includegraphics[scale=0.3]{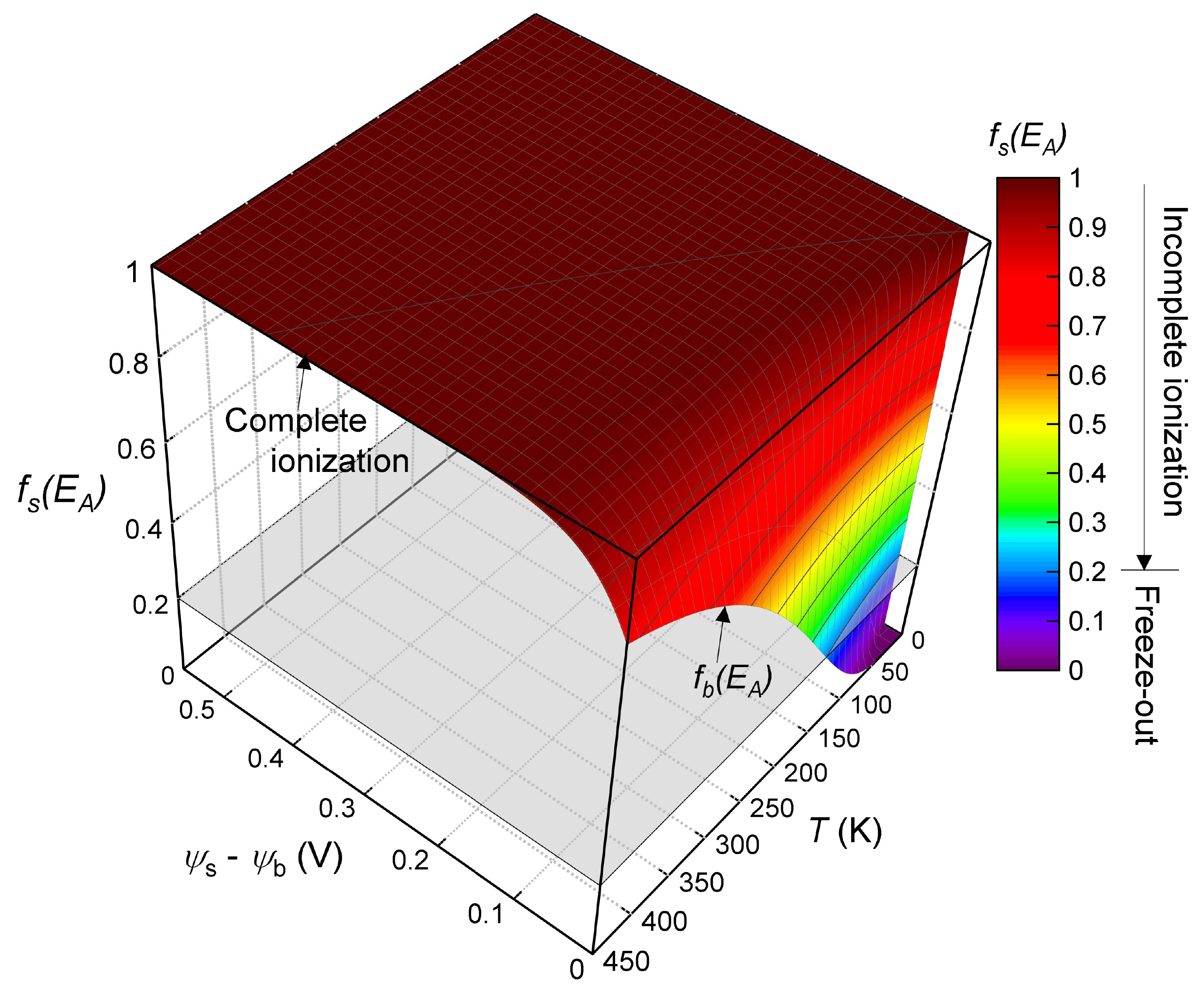}
	\caption{\label{fig:ionization}Dopant ionization at the surface is an interplay between thermal ionization ($T$) and field-assisted ionization ($\psi_s-\psi_b$).  Freeze-out is assumed when 20\,\% of the dopants are ionized. This happens only when $T$ is below $\approx$ \SI{50}{\kelvin} and close to the flatband condition ($\psi_s \approx \psi_b$). When $\psi_s$ increases, a rapid transition takes place to complete ionization for all temperatures. In the flatband condition ($\psi_s = \psi_b$) the ionization probability at the surface is only due to thermal ionization and equals the ionization probability in the bulk, $f_b(E_A)$.}	
\end{figure}
The surface-boundary condition, i.e. the link between $V_{\mathrm{GB}}$ and $\psi_s$, is given by $V_{\mathrm{GB}}=V_{\mathrm{FB}}+\varepsilon_{si}\textbf{E}_s/C_{ox}+(\psi_s-\psi_b)$ where $C_{ox}$ is the oxide capacitance per unit area, and $V_{\mathrm{FB}}$ is the flatband voltage, given by  $V_{\mathrm{FB}}\triangleq\phi_{ms}-Q_{it}/C_{ox}$\cite{tsividis2011operation,sze2006physics}. Here $Q_{it}$ is the interface trap charge density per unit area.
\begin{table*}
	\begin{equation}
	\begin{split}
	Q_m&=-\varepsilon_{si}\sqrt{\frac{2qn_iU_T}{\varepsilon_{si}}\left(e^{\frac{\psi_s-V_{ch}}{U_T}}-e^{\frac{\psi_b-V_{ch}}{U_T}}\right)
		+\frac{2qN_A}{\varepsilon_{si}}\Bigg[\psi_s-\psi_b-U_T\ln\frac{f_s(E_A)}{f_b(E_A)}\Bigg]}+\varepsilon_{si}\sqrt{\frac{2qN_A}{\varepsilon_{si}}\Bigg[\big(\psi_s-\psi_b\big)-U_T\ln\frac{f_s(E_A)}{f_b(E_A)}\Bigg]}
	\end{split}
	\label{qm}
	\end{equation}
\end{table*}
We consider the summation of the discrete acceptor trap energy levels (all donor states are occupied and neutral during turn-on in $n$MOS\cite{sze2006physics}). Each discrete trap-energy-level, $E_{t,j}$, at position $j$ in the bandgap has its particular $\textcolor{black}{N}_{it,j}$-value assigned to it, where $\textcolor{black}{N}_{it}$ is the density-of-interface-traps per unit area. $Q_{it}$ can then be expressed as $Q_{it}=-q\sum_{j}^N\textcolor{black}{N}_{it,j}f_s(E_{t,j})$ 
where $\textcolor{black}{N}$ is the number of interface traps, and 
\begin{equation}\label{fss}
f_s(E_{t,j})=\frac{1}{1+g_te^{\frac{E_{t,j}-E_{F,s}}{kT}}}=\frac{1}{1+g_te^{\frac{\psi_{t,j}-(\psi_s-V_{ch})}{U_T}}}, 
\end{equation}
is the Fermi-Dirac occupation probability of the trap energy level $E_{t,j}$. The RHS of Eq.\,(\ref{fss}) is obtained by defining the trap potentials, $\psi_{t,j}\triangleq(E_{t,j}-E_i)/q$\cite{jazaeri_sallese_2018,yesayan2016charge,jazaeri}. This leads to the flatband voltage 
\begin{figure*}[t]
	\includegraphics[width=\textwidth]{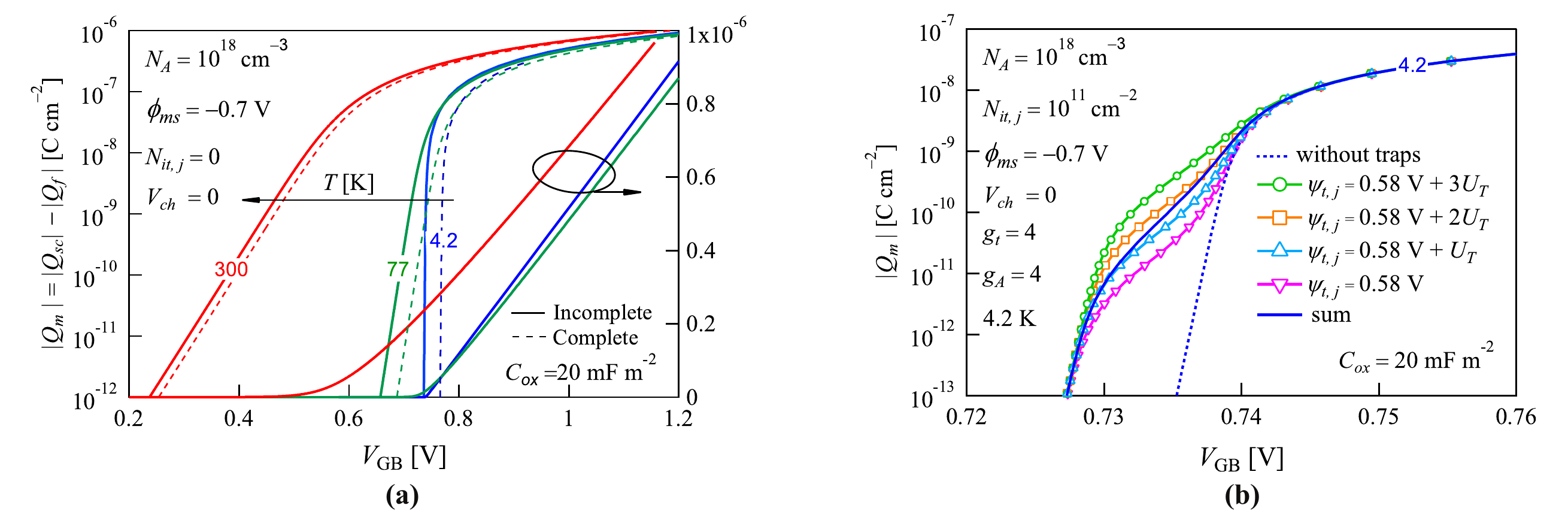}
	\vspace{-0.6cm}
	\caption{\label{fig:Qm} Mobile charge density, (a) without interface traps at RT, \SI{77}{\kelvin} and \SI{4.2}{\kelvin} including incomplete ionization (solid lines), and assuming complete ionization (dashed lines). Incomplete ionization yields a small decrease in the charge-threshold voltage, (b) Influence of four single interface traps close to the conduction band, and their combined effect on the turn-on rate of $Q_m$ at \SI{4.2}{\kelvin}.}	
\end{figure*}
\begin{equation}
V_{\mathrm{FB}}\!=\!\phi_{ms}+\frac{q}{C_{ox}}\sum_j^N \frac{N_{it,j}}{1+g_t\exp\{\left[\psi_{t,j}\!-\!(\psi_s\!-\!V_{ch})\right]/U_T\}}. 
\end{equation}
Plotting $Q_m$ from (\ref{qm}) versus $V_{\mathrm{GB}}$ at 4.2\,K in Fig.\,\ref{fig:Qm}\textcolor{black}{b}, including four interface traps close to the conduction band, reveals how each interface trap degrades the turn-on of $Q_m$ separately, as well as the combined effect of the sum of the interface traps. 
\section{Current derivation}
To derive the current in the linear regime, this core-model assumes drift-diffusion transport, and does not include ballistic nor quantum transport. To verify the drift-diffusion transport-mechanism at cryogenic temperatures, the proposed model for the drain-to-source current will be experimentally validated in Section \ref{sec:exp}. Neglecting the hole-contribution to the current, the expression for the total drain-source current is given by  $I_{\mathrm{DS}}=-\mu_n(W/L)\int_{V_{\mathrm{SB}}}^{V_{\mathrm{DB}}}Q_m(V_{ch})dV_{ch}$, 
where the electron mobility $\mu_n$ is assumed constant along the channel, and $W/L$ is the device aspect-ratio, as indicated in Fig.\,\ref{fig:mos}. In the linear regime, $Q_m$ can be assumed independent of $V_{ch}$. In this case, the total drain-source current is given by $I_{\mathrm{DS}}=-\mu_n(W/L) Q_m V_{\mathrm{DS}}$. In saturation, the integral over $V_{ch}$ cannot be readily solved. Therefore, starting from the drift-diffusion equation gives \begin{equation}\label{driftdiffusion}
I_{\mathrm{DS}}=-\frac{W}{L}\int_{\psi_{s,\mathrm{S}}}^{\psi_{s,\mathrm{D}}}\mu_nQ_md\psi + \frac{W}{L}\int_{Q_{m,\mathrm{S}}}^{Q_{m,\mathrm{D}}}\mu_nU_TdQ_m. 
\end{equation}	
Assuming a linearization of the mobile charge density with respect to the surface potential at constant gate voltage~\cite{sallese} in (\ref{driftdiffusion}), i.e., $Q_m = mC_{ox}(\psi_s-\psi_\mathrm{P})$, with $m\triangleq\partial (Q_m / C_{ox})/\partial\psi_s$ and $\psi_\mathrm{P}$ the pinch-off potential, and integrating, results in an expression for the total drain-source current in saturation, 
\begin{equation}
	I_{\mathrm{DS}}\!=\!\frac{W}{L}\mu_n\left[-\frac{Q_{m,\mathrm{D}}^2-Q_{m,\mathrm{S}}^2}{2mC_{ox}}\!+\!U_T(Q_{m,\mathrm{D}}-Q_{m,\mathrm{S}})\right].
\end{equation} 
$Q_{m,\mathrm{S}}$ and $Q_{m,\mathrm{D}}$ are obtained from (\ref{qm}), setting $V_{ch}$ to zero and $V_{\mathrm{DS}}$ respectively. At cryogenic temperature an improvement in the low-field mobility, $\mu_0$, is observed due to a reduction of the phonon scattering\cite{kirschman1985cold,balestra_physics_2017}. In addition, the mobility reduces at higher gate voltage due to surface roughness scattering at high vertical electric field\cite{balestra_physics_2017}. This mobility reduction can be modeled by $\mu_n=\mu_0/(1+\theta V_{\mathrm{GB}})$ where $\theta$ is the mobility reduction factor.
\section{Experimental Results and discussion}\label{sec:exp}
Room temperature and cryogenic measurements were performed on devices fabricated in a 28-nm bulk CMOS process. The full set of measurements, measurement set-up, and characterization were previously reported in \cite{essderc,jeds}. After measuring at RT, the samples were immersed into liquid helium (4.2\,K) and liquid nitrogen (77\,K) baths with a dipstick. Fig.\,\ref{fig:current}a favorably compares the model with the linear transfer characteristics ($V_{\mathrm{DB}}=\SI{20}{\milli\volt}$) measured at RT and 4.2\,K on a long $n$MOS device with $W/L=$\,3$\upmu$m\,/\,1$\upmu$m, in linear and logarithmic scales. The extracted $\mu_0$-values from the model are in accordance with the characterization performed in\cite{essderc}. Furthermore Fig.\,\ref{fig:current}b analyzes the effect of incomplete ionization, interface traps, and mobility, on the current at 4.2\,K. Note that incomplete ionization reduces the threshold voltage, and interface traps can degrade the subthreshold swing ($SS$) to $\approx$\,10\,mV/decade. The strong increase in the mobility increases the on-state current at 4.2\,K. Fig.\,8 validates the model for the current in saturation ($\vert V_{\mathrm{DB}}\vert=0.9\,\mathrm{V}$) using the measurements performed at RT, 77, and 4.2\,K on a long $p$MOS device with $W/L=$\,3$\upmu$m\,/\,1$\upmu$m, in linear and logarithmic scales. The metal-semiconductor work function difference, $\phi_{ms}$, increases in absolute value at lower temperatures according to the change in $E_F$-position (Fig.\,\ref{fig:extrinsic}). 
\section{\label{appSS}Subthreshold-swing derivation}
\begin{figure*}[t!]
	\includegraphics[width=\textwidth]{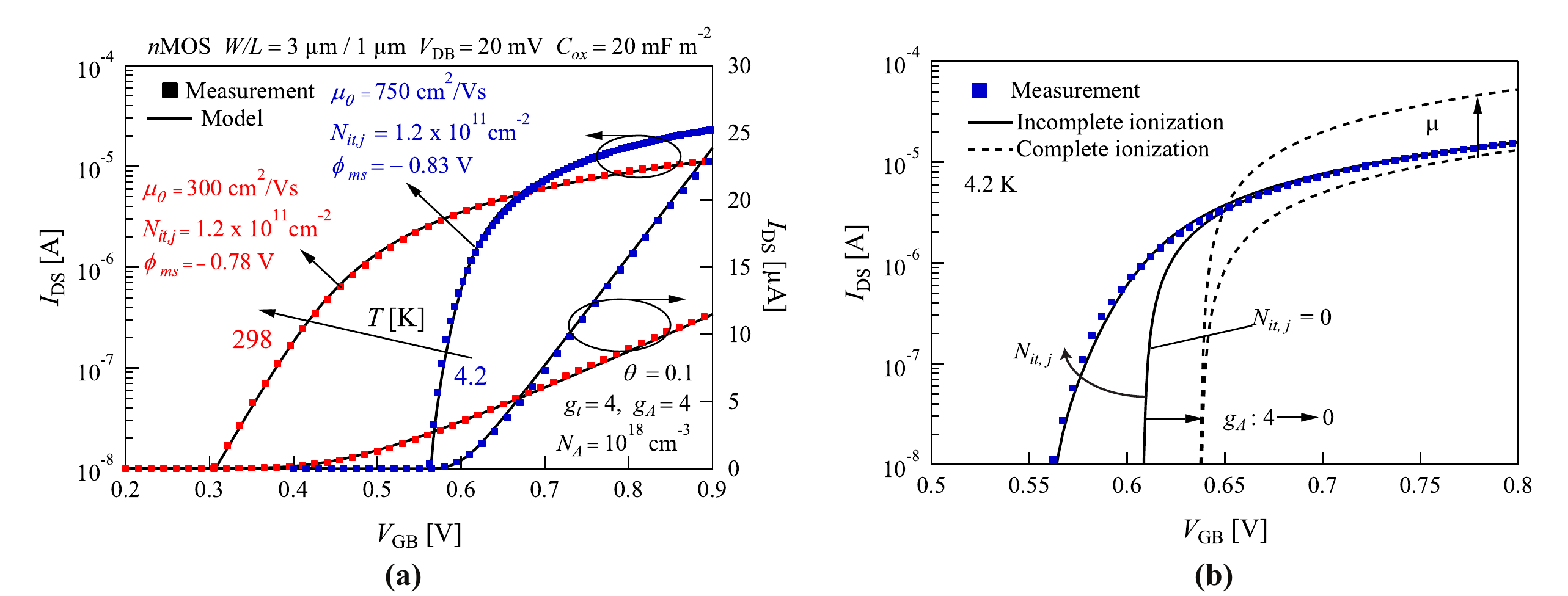}
	\vspace{-0.6cm}
	\caption{\label{fig:current} (a) Linear model validation with measurements at RT and \SI{4.2}{\kelvin} on a long $n$MOS device (28-nm bulk CMOS process). In the measurements the gate voltage was swept from \SI{0.2}{\volt} to \SI{0.9}{\volt} with a step size of \SI{1}{\milli\volt} in order to reliably resolve the steep subthreshold slope at cryogenic temperature. The sets of used physical model parameters are shown. Five interface traps are placed at $\psi_{t,j}=0.58\,\mathrm{V}-2U_T:U_T:0.58\,\mathrm{V}+2U_T$, (b) Overview of the phenomena influencing the current at 4.2\,K: incomplete ionization ($g_A$\,=\,4) leads to a decrease in the threshold voltage; interface traps ($\textcolor{black}{N}_{it,j}$) strongly degrade the subthreshold swing, and mobility ($\mu$) increases the on-state current.}
\end{figure*}

In this section a $SS$-expression including incomplete ionization and temperature-dependent interface trapping is derived. Incomplete ionization is included to prove the minimal influence on $SS$ shown in Fig.\ref{fig:current}b. The temperature-dependency of interface-trap occupation, $f_s(E_t)$, allows to obtain the $\Delta_{SS}$-offset of $\approx$\,10\,mV/dec above the thermal limit, $U_T\ln10$, previously observed on long-channel devices\cite{essderc,jeds}. 

The subthreshold swing, $SS$, is usually expressed as $nU_T\ln10$, where the non-ideality factor or slope factor, $n$, is given by $(\partial V_{\mathrm{GB}}/\partial \psi_s)$, describing the deviation from the thermal limit. Assuming $f_s(E_t)$ from (\ref{fss}) to be one, $(\partial V_{\mathrm{GB}}/\partial \psi_s)$ yields $1+\sqrt{2qN_A\varepsilon_{si}}/\left[C_{ox}(2\sqrt{\psi_s-\psi_b})\right]+q\textcolor{black}{N}_{it}/C_{ox}$\cite{tsividis2011operation,tewksbury_attojoule_1985}. However, at 4.2\,K, and assuming the highest possible doping value below the degenerate limit, a large $\textcolor{black}{N}_{it}$-value in the order of $10^{13}\,\textcolor{black}{\mathrm{cm^{-2}}}$ is extracted to accommodate for a $SS$ of $\approx$\,10\,mV/decade\cite{hafez_assessment_1990,trevisoli_junctionless_2016}, since $\textcolor{black}{N}_{it}$ becomes multiplied with $U_T$ in this expression. However, it should be emphasized that in the used expression for $SS$, the temperature dependency of interface-trap occupation is not taken into account. 
\begin{figure}
	\centering
	\includegraphics[width=0.5\textwidth]{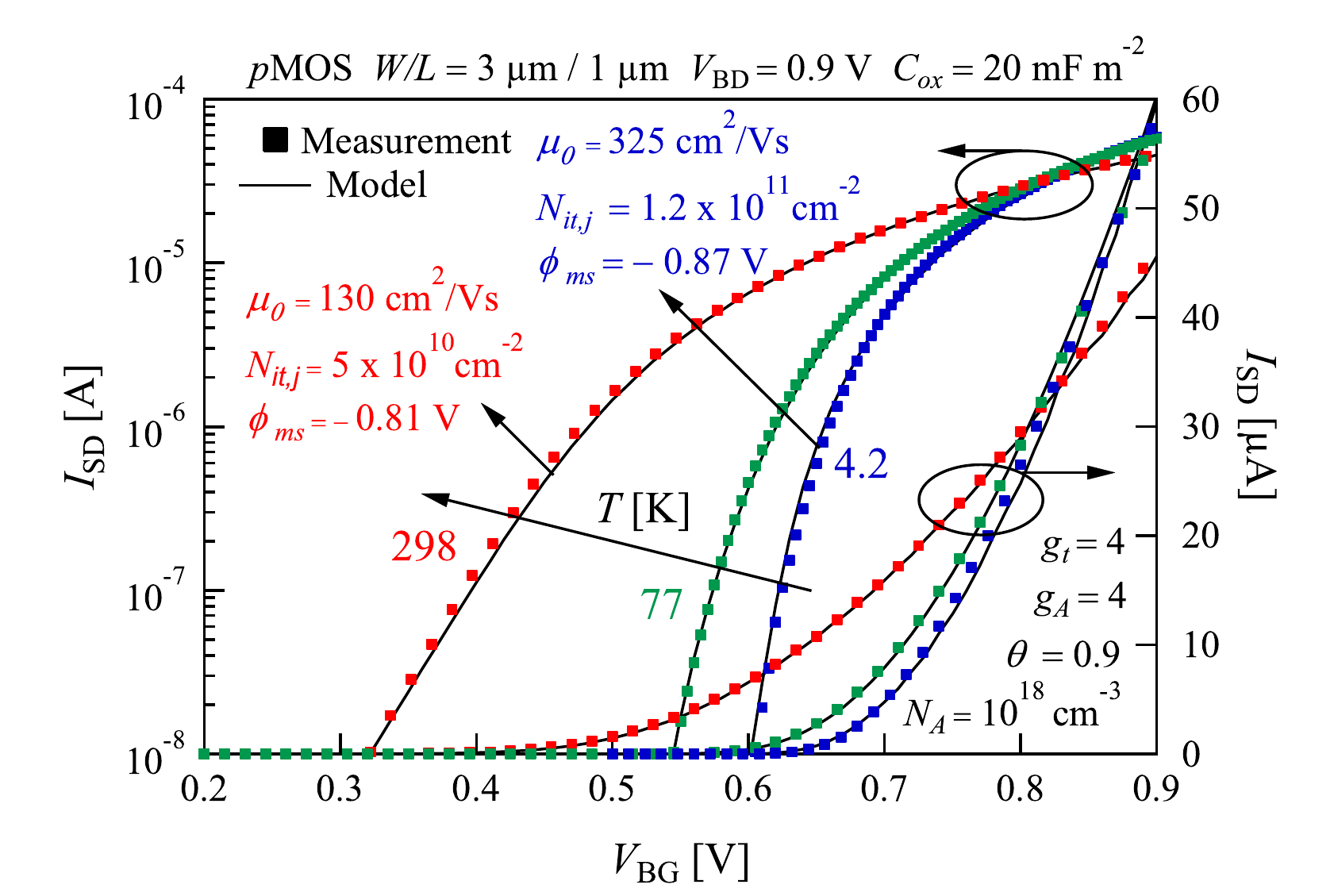}
	\caption{Saturation model validation with measurements at RT, \SI{77}{\kelvin} and \SI{4.2}{\kelvin} on a long $p$MOS device (28-nm bulk CMOS process). Four interface traps are placed at $\psi_{t,j}=0.58\,\mathrm{V}-2U_T:U_T:0.58\,\mathrm{V}+U_T$. The used physical model parameters for RT and \SI{4.2}{\kelvin} are shown in the figure. For \SI{77}{\kelvin}, the model parameters are: $\phi_{ms}=-0.87$\,V, $N_{it,j}=\SI{1.1e11}{\per\centi\meter\squared}$, and $\mu_0=\SI{300}{\centi\meter\squared\per\volt\per\second}$.}
	\label{fig:saturation}
\end{figure}
Relying on drift-diffusion transport  in the linear regime and assuming $\mu$ independent of $V_{\mathrm{GB}}$, the subthreshold slope, $SS^{-1}$, is given by
\begin{equation}\label{slope}
SS^{-1}=\frac{1}{\ln10}\frac{1}{Q_m}\frac{\partial Q_m}{\partial \psi_s}\frac{\partial \psi_s}{\partial V_{\mathrm{GB}}}.
\end{equation}
The factor $(1/{Q_m})(\partial Q_m/\partial \psi_s)$ is found from (\ref{qm}) by considering $Q_m \ll Q_f$ or $Q_{sc} \approx Q_f$ in the subthreshold region. After some mathematical manipulation, we find 
\begin{equation}
\begin{split}
\frac{1}{Q_m}\frac{\partial Q_m}{\partial \psi_s}=q\varepsilon_{si}&\Bigg[\frac{1}{Q_mQ_f}n_ie^{\frac{\psi_s-V_{ch}}{U_T}}\\&-\frac{1}{Q_f^2}N_A\left(1-\frac{U_T}{f(E_A)}\frac{\partial f(E_A)}{\partial \psi_s}\right)\Bigg]. 
\end{split}
\label{qmdqm}
\end{equation}
Merging (\ref{qf}) and (\ref{qmdqm}), and inverting, gives
\begin{eqnarray}
SS= \frac{2N_A\ln10\Big[\big(\psi_s-\psi_b\big)-U_T\ln\frac{f(E_A)}{f_b(E_A)}\Big]}{\frac{Q_f}{Q_m}\underbrace{n_ie^{\frac{\psi_s-V_{ch}}{U_T}}}_{\text{(a)}}-N_A\Big(1-\frac{U_T}{f(E_A)}\frac{\partial f(E_A)}{\partial \psi_s}\Big)}\frac{\partial V_{\mathrm{GB}}}{\partial \psi_s}.
\label{sss}
\end{eqnarray}
The following relation can be derived for (a) in the above equation (see Appendix),  
\begin{eqnarray}
n_ie^{\frac{\psi_s-V_{ch}}{U_T}}=\frac{Q_m}{Q_f} \frac{2N_A\Big[\big(\psi_s-\psi_b\big)-U_T\ln\frac{f(E_A)}{f_b(E_A)}\Big]}{U_T}.
\label{meq}
\end{eqnarray}
Plugging this in (\ref{sss}) one finds 
\begin{eqnarray}
SS=U_T\ln(10) \frac{1}{1-\frac{U_T\Big(1-\frac{U_T}{f_s(E_A)}\frac{\partial f_s(E_A)}{\partial \psi_s}\Big)}{2\Big[\big(\psi_s-\psi_b\big)-U_T\ln\frac{f_s(E_A)}{f_b(E_A)}\Big]}}\frac{\partial V_{\mathrm{GB}}}{\partial \psi_s},
\end{eqnarray} 
where  
\begin{equation}
\begin{split}
\frac{\partial V_{\mathrm{GB}}}{\partial \psi_s}=1&+\frac{\sqrt{2qN_A\varepsilon_{si}}}{C_{ox}}\frac{1-\frac{U_T}{f_s(E_A)}\frac{\partial f_s(E_A)}{\partial \psi_s}}{2\sqrt{(\psi_s-\psi_b)-U_T\ln\frac{f_s(E_A)}{f_b(E_A)}}}\\
&-\frac{q}{C_{ox}}\sum_j\textcolor{black}{N}_{it,j}\frac{\partial f_s(E_{t,j})}{\partial \psi_s}. 
\end{split}
\end{equation}
follows from the surface-boundary condition derived in Sec.\,\ref{sec:traps}. In the subthreshold region, far above the flatband condition, $f_s(E_A)=1$ can be assumed (Fig.\,\ref{fig:ionization}), and $U_T \ll 2(\psi_s-\psi_b)$, leading to $SS=U_T\ln10(\partial V_{\mathrm{GB}}/\partial{\psi_s})$ with 
\begin{equation}
\begin{split}
\frac{\partial V_{\mathrm{GB}}}{\partial \psi_s}=1&+\frac{\sqrt{2qN_A\varepsilon_{si}}}{C_{ox}}\frac{1}{2\sqrt{\psi_s-\psi_b}}\\&-\frac{q}{C_{ox}}\sum_j\textcolor{black}{N}_{it,j}\frac{\partial f_s(E_{t,j})}{\partial \psi_s}. 
\end{split}
\label{Vgbpsis}
\end{equation}
Taking the derivative of (\ref{fss}), (\ref{Vgbpsis}) becomes
\begin{equation}
\begin{split}
&\frac{\partial V_{\mathrm{GB}}}{\partial \psi_s}=1+\frac{\sqrt{2qN_A\varepsilon_{si}}}{C_{ox}}\frac{1}{2\sqrt{\psi_s-\psi_b}}\\&+\frac{q}{C_{ox}}\frac{1}{U_T}\sum_j\textcolor{black}{N}_{it,j}\frac{g_t\exp[(\psi_{t,j}-\psi_s)/U_T]}{\{1+g_t\exp[(\psi_{t,j}-\psi_s)/U_T]\}^2}. 
\end{split}
\label{Vgbfs}
\end{equation}
Note the appearance of a factor $1/U_T$ in the third term on the RHS of (\ref{Vgbfs}). Placing a discrete interface trap, $\psi_{t,j}$, at each $\psi_s$-value in the subthreshold region, and assuming a uniform $\textcolor{black}{N}_{it,j}$-value for each trap, leads to
\begin{equation}
\begin{split}
\frac{\partial V_{\mathrm{GB}}}{\partial \psi_s}=1+\frac{\sqrt{2qN_A\varepsilon_{si}}}{C_{ox}}\frac{1}{2\sqrt{\psi_s-\psi_b}}+\frac{q\textcolor{black}{N}_{it}}{C_{ox}}\frac{1}{U_T}\frac{g_t}{(1+g_t)^2}. 
\end{split}
\label{Vgbfss}
\end{equation}
The first two terms in (\ref{Vgbfss}) yield the non-ideality or slope factor without interface traps, $n_0$. The $SS$-expression becomes 
\begin{equation}
\begin{split}
SS=n_0U_T\ln10+\frac{q\textcolor{black}{N}_{it}}{C_{ox}}\frac{g_t}{(1+g_t)^2}\ln10
\end{split}
\label{Vgbfsss}
\end{equation}
where the second term on the RHS is the sought $\Delta_{SS}$-offset. The non-ideality factor $n_0$ has an upper bound of 2 mainly related to doping. Therefore, at 4.2\,K, the first term on the RHS of (\ref{Vgbfsss}) is limited by 1.6\,mV/decade. Note that $\textcolor{black}{N}_{it}$ does not become multiplied with $U_T$ in (\ref{Vgbfsss}). Therefore, assuming a reasonable value for $\textcolor{black}{N}_{it}=3\times 10^{11}\,\textcolor{black}{\mathrm{cm}^{-2}}$, the second term gives $\approx$ 9\,mV/decade (with $C_{ox}=\SI{20}{\milli\farad\per\square\meter}$ and $g_t$\,$=$4). Together they yield the $SS$-degradation observed on a long $n$MOS device at 4.2\,K. At \SI{77}{\kelvin}, a similar calculation using $n_0=1.08$ and $N_{it}=3\times 10^{11}\,\mathrm{cm}^{-2}$ gives 25\,mV/dec, corresponding to the subthreshold swing measured \cite{essderc} on a long $p$MOS device at \SI{77}{\kelvin} (Fig.\,\ref{fig:saturation}).

\section{\label{sec:concl}Conclusion}
A theoretical MOS transistor model is developed valid from room temperature down to liquid-helium temperature. The model relies on Boltzmann statistics,  verified in the limit to zero Kelvin, and includes incomplete ionization, interface traps, bandgap temperature dependency, and mobility reduction. It is evidenced that incomplete ionization maintains the non-degeneracy of a semiconductor at deep-cryogenic temperatures, and leads to a decrease in the threshold voltage on top of the overall increase due to Fermi-Dirac distribution scaling. The Fermi-Dirac temperature-dependency of interface-trap occupation degrades the subthreshold swing down to 4.2\,K. An expression for the subthreshold swing including incomplete ionization and temperature-dependent interface trapping is derived. The proposed model builds the indispensable physical foundation for future low-temperature CMOS circuit design.
\appendix
\section{Formula}
Starting from $Q_m+Q_f=Q_{sc}$, we can write $(Q_m+Q_f)^2=\varepsilon_{si}^2\textbf{E}_s^2$, with $\textbf{E}_s$ given by (\ref{es}). Solving a quadratic equation for $Q_m$ leads to 
\begin{equation}
\frac{Q_m}{Q_f}=-1+\sqrt{1+\frac{2qn_iU_T\varepsilon_{si}}{Q_f^2}e^{\frac{\psi_s-V_{ch}}{U_T}}}
\end{equation}
where we neglected the exponential term in $\psi_b/U_T$. In the subthreshold region ($Q_m\ll Q_f$), $\sqrt{1+x}$ can be approximated by $1+x/2$ for $x\rightarrow0$. Using (\ref{qf}) for $Q_f^2$ leads to (\ref{meq}). 
\appendices

\bibliographystyle{IEEEtran}
\bibliography{ted}

\end{document}